\documentclass[preprint,showpacs,eqsecnum,amsmath,amssymb,aps,prd,nofootinbib]{revtex4}
\usepackage{graphicx}
\newcommand{\rf}[1]{(\ref{#1})}
\newcommand{\hc}{\ensuremath{\mbox{h.c.}} }

\begin{document}

\preprint{
\begin{tabular}{r}
FTUV--02--1230\\ 
IFIC/02--65 
\end{tabular}
}

\title{Universal extra dimensions and $Z \rightarrow b \bar{b}$ }

\author{J.F. Oliver, J. Papavassiliou, and A. Santamaria}
\affiliation{Departament de F\'{\i}sica Te\`orica 
and IFIC, Universitat de Val\`encia -- CSIC\\
E-46100 Burjassot (Val\`encia), Spain}

\date{\today}

\begin{abstract}
We study,  at the  one loop level,  the dominant contributions  from a
single  universal  extra  dimension  to the  process  \(  Z\rightarrow
b\bar{b} \).   By resorting to the  gaugeless limit of  the theory we
explain why the  result is expected to display  a strong dependence on
the mass of the top-quark,  not identified in the early literature.  A
detailed calculation corroborates this  expectation, giving rise to a
lower bound for the compactification scale which is comparable to that
obtained  from the $\rho$  parameter.  An  estimate of  the subleading
corrections is  furnished, together  with a qualitative  discussion on
the  difference   between  the  present  results   and  those  derived
previously for the non-universal case.
\end{abstract}

\maketitle

\section{Introduction}

Models with large extra dimensions 
\cite{Arkani-Hamed:1998nn,Arkani-Hamed:1998rs,Antoniadis:1990ew,
Antoniadis:1994jp} 
have been extensively studied in recent years, and 
have served as a major source of inspiration in  
the ongoing search of physics beyond the 
Standard  Model (SM). The general idea behind these scenarios  
is that the ordinary four  dimensional SM 
emerges as the low
energy effective theory  of more fundamental
models living in five  or more dimensions
with the extra  dimensions compactified.
The
effects  of  the  extra   dimensions  are  communicated  to  the  four
dimensional world through the presence of infinite towers of Kaluza-Klein (KK) 
modes, which modify qualitatively the  behavior of the low-energy theory.  In
particular,  the non-renormalizability  of  the theory  is found  when
summing  the infinite  tower of  KK states.   The size  of  the extra
dimensions  can be  surprisingly large  without  contradicting present
experimental          data          (see         for          instance
\cite{Antoniadis:1994yi,Pomarol:1998sd,
Antoniadis:1999bq,Nath:1999mw,Masip:1999mk,
Delgado:1999sv,Rizzo:1999br,Carone:1999nz,Nath:1999fs,Muck:2001yv,Muck:2002er}).   
This  offers
the  exciting possibility of  testing  these models  in the  near
future, since the lowest  KK states,  if  light enough,
could be produced in the next generation of accelerators.

Extra dimensions may  or may
not be accessible  to all known fields, depending  on the specifics of
the underlying, more fundamental theory.  
Scenarios  where all  SM
fields  live   in  higher  dimensions  have been 
the focal point of particular 
attention~\cite{Carone:1999nz,Appelquist:2000nn}).
This type  of  extra
dimensions is refered to in the literature as 
``universal extra dimensions'' (UED).
From the  phenomenological point  of view,
the most  characteristic feature of such theories  is the conservation
of the KK  number at each elementary interaction  vertex 
\cite{Carone:1999nz,Appelquist:2000nn}.
As a result,
and contrary to  what happens in the non-universal  case, the coupling
of any excited (massive) KK mode to two zero modes is prohibited.  
This fact alters profoundly their production mechanisms:
using normal  (zero-mode)  particles as  initial
sates, such modes cannot be resonantly produced, nor 
can a single KK mode appear in the final states, but instead they  
must be   pair-produced. 
In addition, the conservation of the KK number  
leads to the appearance of heavy stable
(charged   and  neutral)  particles,   which  seem to pose  cosmological
complications (e.g. nucleosynthesis) \cite{Appelquist:2000nn}; 
however, one-loop effects may overcome such problems 
\cite{Servant:2002aq}. 
Finally, this conservation yields the additional 
important feature that, 
the constraints on the size of the extra dimensions 
which are obtained from SM precision measurements are 
less stringent; this is so because the extra modes 
do not affect 
the tree-level predictions, and make their 
presence felt only through loop corrections. 
This last point merits particular attention, given 
its phenomenological importance, together with the fact that  
loop calculations in the context of such 
theories constitute a relatively unexplored territory. 

In general the precision electroweak observables 
most sensitive to radiative corrections, whether 
from within the SM or from its extensions, 
are those
enhanced by the large top-quark mass : $R_b$, or equivalently, 
the process
\( Z\rightarrow b\bar{b} \)
\cite{Akhundov:1986fc,Bernabeu:1988me,Bernabeu:1991ws,Beenakker:1988pv},
the \( B-\bar{B} \) -mixing \cite{Buchalla:1996vs}, 
and the \( \rho  \) parameter. 
These observables have already been considered in models with extra dimensions.
Thus, $R_b$ was considered, for instance, in 
\cite{Papavassiliou:2000pq,Appelquist:2000nn,DelAguila:2001pu}, 
\( B-\bar{B} \) was considered first in~\cite{Papavassiliou:2000pq} and,
recently, it has been studied in the context of UED
in~\cite{Chakraverty:2002qk,Buras:2002ej},
In the case of theories with UED the study of the 
corrections to the \( \rho  \) parameter has yielded 
a lower bound on the size 
of the compactification scale, the inverse of the compactification radious,
$R$,
of about 300 GeV~\cite{Appelquist:2000nn}. In this paper we will study 
in detail the bound obtained on the size of a single UED from the 
process \( Z\rightarrow b\bar{b} \). 
Our experience 
with the radiative corrections induced by 
the SM particles suggests 
that the bounds obtained 
from this process could in principle be 
comparable to those extracted from 
the $\rho$ parameter; the reason is that 
the aforementioned enhancement 
induced by the dependence on the mass of the top-quark takes 
place in both cases. 
A simple one-loop calculation, motivated by the behavior of the theory 
in its gaugeless limit and subsequently corroborated by 
a more detailed analysis, 
reveals that,   
contrary to what has been claimed in
\cite{Appelquist:2000nn},
the leading corrections to the left-handed $Zb\bar{b}$ coupling, 
$g_L$, due to the KK modes corresponding to a single UED,
display a strong dependence on the mass of the 
top-quark (they are proportional to $m_t^4$). 
This 
fact makes the bounds obtained from $R_b$ 
comparable to those obtained from the $\rho$ parameter; 
in particular we find $R^{-1} > 300$~GeV at 95\%~CL.

The paper is organized as follows: In section II we 
start from the five-dimensional Lagrangian and after standard 
manipulations we derive the 
corresponding four-dimensional interactions and mass spectrum, 
paying particular attention to the interactions involving top 
and bottom quarks. 
In section III we first discuss 
the physical arguments which suggest a strong dependence of the 
result on the top-quark mass;  
then we present a more detailed 
one-loop calculation which demonstrates precisely the 
announced leading behavior. The subleading corrections, e.g.     
terms suppressed by an additional factor ${\cal O}(M_W^2/m_t^2)$ are also 
estimated. Finally, in section IV we present our conclusions. 

\section{The Lagrangian}

We will concentrate on 
the electroweak part, 
$SU(2)_L\times U(1)_Y$, of the SM Lagrangian, written in five 
dimensions; we will 
denote by $x$ the four normal coordinates, and by $y\equiv x^4$ the fifth one, 
which will undergo compactification.

The Lagrangian 
$\mathcal{L}$ assumes the form 
\begin{equation}
\label{eq:primera}
\mathcal{L}=\int_0^{\pi R} dy (\mathcal{L}_A + \mathcal{L}_{H} +
\mathcal{L}_{F} + \mathcal{L}_Y) \,,
\end{equation}
where 
\begin{eqnarray}
\mathcal{L}_A &=&-\frac{1}{4}F^{MNa}F_{MN}^a
 -\frac{1}{4}F^{MN}F_{MN} ~,\nonumber\\
\mathcal{L}_H &=& (D_M \Phi)^\dagger D^M \Phi -V(\Phi)~,\nonumber\\
\mathcal{L}_F &=& \overline{Q}(i\Gamma^M D_M)Q + \overline{U}(i\Gamma^M
D_M)U+\overline{D}(i\Gamma^M D_M)D~, \nonumber\\
\mathcal{L}_Y &=& -\overline{Q} \widetilde{Y}_u \Phi^c  U - \overline{Q}
\widetilde{Y}_d \Phi D + \hc
\end{eqnarray}
In the above formulas   
$M,N=0,1,2,3,4$ are the five-dimensional Lorentz indices,  
$F_{MN}^a = \partial_M W_N^a - \partial_N W_M^a+g \epsilon^{abc}W_M^b
W_N^c$ 
is the field
strength associated with the $SU(2)_L$ gauge group, 
and $F_{MN}=\partial_M B_N - \partial_N B_M$
that of the $U(1)_Y$ group.
The covariant derivative is defined as
$D_M\equiv\partial_M-i\widetilde{g}W_M^a T^a-i \widetilde{g}^\prime
B_M Y$, where  
$\widetilde{g}$ and $\widetilde{g}^\prime$ are the 
five-dimensional
gauge coupling
constants of $SU(2)_L$ and $U(1)_Y$, respectively, 
and $T^a$ and $Y$ are the corresponding generators.
$\Gamma_M$ denote the five dimensional gamma matrices, $\Gamma_\mu
= \gamma_\mu$ and $\Gamma_4 = i \gamma_5$, and  
the metric convention is
$g_{MN}=(+,-,-,-,-)$. The fermionic fields $Q$, $D$ and $U$ are
four-component spinors and carry the same quantum numbers as the corresponding 
SM fields. $SU(2)$ and color indices have been suppressed. 
Finally, $\Phi$ and $\Phi^c=i\tau^2 \Phi^\ast$ denote 
the standard Higgs doublet and its charge conjugated field, and
$\widetilde{Y}_u$ are the Yukawa matrices in the five dimensional
theory; they mix different generations, whose indices are suppressed.  
We do not include lepton or gluon couplings because they are not relevant
for our discussion.

Next, as usual, we assume that 
the fifth dimension
is compactified on a circle of radius \( R \) 
in which the points $y$
and $-y$ are identified (e.g. an orbifold \( S^{1}/{\mathbb Z}_{2} \)).
Fields even under the \( {\mathbb Z}_{2} \) symmetry 
will have zero modes
and will be present in the low energy theory. 
Fields odd under \( {\mathbb Z}_{2} \)
will only have KK modes and will disappear from the low energy spectrum. One
chooses the Higgs doublet to be even under the \( {\mathbb Z}_{2} \) symmetry
in order to have a standard zero mode Higgs field.
Then we carry out the Fourier expansion of the fields,
\begin{eqnarray}
A_\mu(x,y) & = &  \frac{1}{\sqrt{\pi R}}A_\mu^{(0)}(x) +
\frac{\sqrt{2}}{\sqrt{\pi R}}\sum_{n=1}^{\infty} A_\mu^{(n)}(x) 
\cos\left(\frac{n y}{R}\right)~,\nonumber\\
A_5 (x,y) & = &\frac{\sqrt{2}}{\sqrt{\pi R}}\sum_{n=1}^{\infty} A_5^{(n)}(x)
\sin\left( \frac{n y}{R}\right)~,\nonumber\\ 
Q (x,y) &=& \frac{1}{\sqrt{\pi R}} Q_L^{(0)}(x)
  + \frac{\sqrt{2}}{\sqrt{\pi R}} \sum_{n=1}^{\infty} \left[ Q_L^{(n)}(x)
\cos\left( \frac{n y}{R}\right) + Q_R^{(n)}(x) \sin\left( \frac{n
y}{R}\right)\right]~,\nonumber\\ 
U (x,y) & = & \frac{1}{\sqrt{\pi R}} U_R^{(0)}(x) 
 + 
\frac{\sqrt{2}}{\sqrt{\pi R}} \sum_{n=1}^{\infty} \left[ U_R^{(n)}(x)
\cos\left( \frac{n y}{R}\right) + U_L^{(n)}(x) \sin\left( \frac{n
y}{R}\right)\right]~,
\label{eq:Fourier}
\end{eqnarray}
where the expansion for $A_\mu$ applies to any of the gauge fields 
and (after suppressing the Lorentz index $\mu$) for the Higgs doublet, 
whereas that of $A_5$ applies 
to the fifth component of the gauge fields. Similarly, the
expansion for $U$ is valid also for $D$. The above 
expansions allow us to
carry out the standard $y$ integration in Eq.~\rf{eq:primera}, and 
obtain the KK spectrum and the relevant interaction terms. 
We will mainly be interested in third generation quarks, thus,
$Q^{(n)}_t$ and $Q^{(n)}_b$ will refer to the
upper and lower parts of the doublet $Q$ and the $U^{(n)}$ will be 
KK modes of right-handed top quarks. 
In particular, 
the relation between the mass- and gauge-eigenstates of the 
KK quarks can be expressed as  
\begin{equation}
\label{eq:chbasis}
\left[
\begin{array}{c}
Q^{(n)}_t\\
U^{(n)}
\end{array}\right]
= 
\left[\begin{array}{cc}
~\gamma_5 \cos(\alpha_n^t) & \sin(\alpha_n^t) \\
- \gamma_5 \sin(\alpha_n^t) & \cos(\alpha_n^t)
\end{array}\right]
\left[\begin{array}{c}
Q^{\prime (n)}_t\\
U^{\prime(n)}
\end{array}\right]~,
\end{equation}
and the mixing angle is given by $\tan(2\alpha_n^t)= m_t/m_n$, 
where $m_n \equiv n/R$.  
The case of $Q^{(n)}_b$ is similar but since we are neglecting all
mass scales except $m_t$ and $m_n$ the mass eigenstate is simply 
$Q^{\prime (n)}_b= \gamma_5 Q^{ (n)}_b$. 
The mass spectrum assumes the form (we remove the primes)
\begin{equation}
\label{eq:CUEDspectrum}
m_{Q^{n}_b}= m_n\,,  \,\,\,\,\,\,
m_{Q^{n}_t}=m_{U^{n}}= \sqrt{m_t^2+m_n^2}\,.
\end{equation}

The couplings between the quarks and the scalar 
modes are important to our purposes, 
because they are proportional to $m_t$. 
In contrast to what
happens within the SM, they will be physical degrees of freedom,
i.e. they cannot be gauged away by choosing, for example, a 
unitary-type of gauge.
After dimensional reduction,
the fifth components of the charged 
gauge fields, $W_5^{-(n)}$,  mix with the KK modes of 
the charged component $\Phi^{-(n)}$
of the Higgs doublet. After diagonalization one obtains  
a physical boson, $\Phi_P^{-(n)}$, and a Goldstone boson $\Phi_G^{-(n)}$ that
will contribute to the mass of the KK gauge bosons. In particular,
\begin{subequations}
\begin{eqnarray}
\Phi_G^{-(n)} & = & \frac{m_n W_5^{-(n)}+i M_W \Phi^{-(n)}}{\sqrt{m_n^2 +
M_W^2}} \stackrel{\stackrel{n\not=0}{M_W\to 0}}{\longrightarrow} W_5^{-(n)}~, \\
\Phi_P^{-(n)} & = & \frac{i M_W W_5^{-(n)} + m_n \Phi^{-(n)}}{\sqrt{m_n^2 +
M_W^2}} \stackrel{\stackrel{n\not=0}{M_W\to 0}}{\longrightarrow} \Phi^{-(n)}
\label{eq:physicalHiggs}~.
\end{eqnarray}
\end{subequations}
As seen from the expansion, the $W_5$ has no zero mode, there is no
physical zero mode $\Phi_P^{-(0)}$ and the zero mode Goldstone boson
comes entirely from the zero mode Higgs field. On the other hand, for
$1/R \gg M_W$ the KK Goldstone bosons are mainly the $W_5^{-(n)}$, while the
physical scalars are mainly the KK modes of the Higgs doublet $\Phi^{-(n)}$.
Their couplings are exactly the same as those of the Goldstone bosons
of the SM, e.g.
\begin{equation}
\label{eq:higgscouplings}
\mathcal{L}_Y = \frac{\sqrt{2}}{v} m_t V_{tj} \overline{U}^{(n)}_{R}
Q^{(0)}_{j\;L} \Phi^{+(n)}+ \hc~,
\end{equation}
where we have written only the  third quark generation and in the following
we will neglect the mixings, $V_{tj}\approx \delta_{tj}$.

The electroweak symmetry breaking proceeds  
by minimizing a Higgs potential of the standard form, e.g. 
$V(\Phi)=-\mu^2\Phi^\dagger \Phi+ 
\widetilde{\lambda} (\Phi^\dagger \Phi)^2$.
The mass terms of the different KK scalar modes are given by
$m^2_{\Phi^n}=-\mu^2+m_n^2$, in such a way that if $\mu < R^{-1}$ 
only the neutral component of the fundamental mode,
$\Phi^{(0)}_0$, gets a vacuum expectation value (VEV),
$\langle \Phi^{(0)}_0\rangle=v/\sqrt{2}$. At low energy, when no KK modes
can be produced, and at tree-level this model coincides exactly with the SM. 
In particular, the VEV of the zero mode Higgs doublet induces
mixing between $W_{\mu 3}^{(0)}$ and $B_{\mu}^{(0)}$ giving rise
to a massless photon, $A_\mu^{(0)}$, and a massive 
$Z$ boson, $Z_\mu^{(0)}$.
 
After a bit of algebra one arrives to the expression of the couplings
of the $Z$ boson with the KK modes of the rest of the fields, given by 
\begin{equation}
\label{eq:couplingZ}
\mathcal{L}_Z =
\frac{g}{2c_w}Z_\mu^{(0)}[J^{\mu(0)} + J^{\mu(n)} + J^{\mu(n)}_{\Phi}]~,
\end{equation}
where the $J^{\mu(0)}$ is the usual SM neutral current, and 
\begin{eqnarray}
J^{\mu(n)} &=& \left(1-\frac{4}{3}
s_w^2\right){\overline{Q}^{(n)}_t} \gamma^\mu Q^{(n)}_t
 - \left(1-\frac{2}{3}
s_w^2\right){\overline{Q}^{(n)}_b}\gamma^\mu Q^{(n)}_b
   - \left(\frac{4}{3}
s_w^2\right){\overline{U}^{(n)}}\gamma^\mu U^{(n)} + \ldots ~,\nonumber\\
J^{\mu(n)}_{\Phi} &=&  (-1 +2s_w^2) \Phi^{+(n)}i\partial^\mu\Phi^{-(n)}
+\hc~.
\label{eq:sourcen}
\end{eqnarray}
Here $Q_t^{(n)}$, $Q_b^{(n)}$ and $U^{(n)}$ are Dirac spinors, and 
the ellipses denote the contribution of $D^{(n)}$ fields, which are not
relevant for our calculation.
Similarly, the interaction of the charged bosons with the quarks 
is given by 
\begin{equation}
\mathcal{L}_W = \frac{g}{\sqrt{2}}[\bar{b}_L\gamma^\mu
  W_\mu^{-(n)}Q_{tL}^{(n)} -i\bar{b}_L W_5^{-(n)}Q_{tR}^{(n)} +
  h.c.]~.
\end{equation}
The couplings of the photon may be derived
similarly. 
From the above equations it
is straightforward to extract the necessary Feynman rules for
our calculations.  
The couplings of the photon may be derived
similarly.

\section{Calculating $Z \rightarrow b \bar{b}$ }

In this section we will compute the corrections to the 
effective \( Zb\bar{b} \) coupling due to the presence of 
a single UED. Shifts in the  \( Zb\bar{b} \) coupling due to 
radiative corrections, either from within the SM or from 
new physics,
affect observables such as the branching
ratio 
$R_{b}=\Gamma _{b}/\Gamma _{h}$,
where $\Gamma _{b}=\Gamma (Z\rightarrow b\bar{b})$
and $ \Gamma _{h}=\Gamma (Z\rightarrow \mathrm{hadrons})$,
or the left right
asymmetry $A_b$. These type of corrections can be   
treated  uniformly by expressing 
them as a modification to the tree level couplings $g_{L(R)}$
defined as
\begin{equation}
\label{eq:definitions}
\frac{g}{c_W}\overline{b} \gamma^\mu(g_L P_L +g_R P_R)b Z_\mu~.
\end{equation}
$Z$ and $b$'s are SM fields, $P_{L(R)}$ are the chirality projectors 
and 
\begin{subequations}
\begin{eqnarray}
g_L& = & -\frac{1}{2}+\frac{1}{3}s_W^2+\delta g_L^{\mathrm{SM}}
+\delta g_L^{\mathrm{NP}}~,\\
g_R & = &\frac{1}{3}s_W^2 + \delta g_R^{\mathrm{SM}} 
+\delta g_R^{\mathrm{NP}}~,
\end{eqnarray}
\end{subequations}
where we have separated radiative corrections coming from SM
contributions and from new physics, (NP). 
It turns out that, both within the SM as well as in most of its 
extensions,  only $g_L$
receives corrections proportional to $m_t^2$ at the one loop level,
due to the
difference in the couplings between the two chiralities. 
In particular, a shift $\delta g_L^{NP}$ in the value of $g_L$
 due to new physics  
translates into a shift in $R_{b}$
given by
\begin{equation}
\label{eq:rbab}
\delta R_b = 2 R_b(1-R_b)\frac{g_L}{g_L^2+g_R^2} \delta g_L^{\mathrm{NP}}~, 
\end{equation}
and to a shift in the left-right asymmetry $A_b$ given by
\begin{equation}
\label{eq:asym}
\delta A_b = \frac{4 g_R^2 g_L}{(g_L^2+g_R^2)^2} \delta g_L^{\mathrm{NP}}~.
\end{equation}
These equations, when compared with experimental data, will be used to set
bounds on the compactification scale.

By far the
easiest way to compute the leading top-quark-mass dependent one-loop 
corrections to $\delta g_L$ from the SM itself, 
$\delta g_L^{\mathrm{SM}}$, 
is to resort to the gaugeless
limit of the SM~\cite{Lytel:1980zh}, e.g. the limit where the gauge couplings $g$ and
$g^\prime$, corresponding to the gauge groups $SU(2)_L$ and $U(1)_Y$
respectively, are switched off. In that limit the gauge bosons play
the role of external sources and the only propagating fields are the
quarks, the Higgs field, and the charged and neutral Goldstone bosons
$G^{\pm}$ and $G^0$. As explained in
\cite{Barbieri:1993dq,Barbieri:1992nz} one may relate the 
one-loop vertex $Z b \bar{b}$ to the corresponding $G^0 b \bar{b}$ vertex by
means of a Ward identity; the latter is a direct consequence of
current conservation, which holds for the neutral current before and
after the Higgs doublet acquires a vacuum expectation value $v$. 

\begin{figure}[ht]
\includegraphics[scale=0.6]{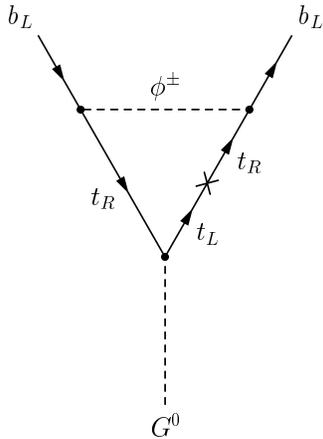}
\caption{\label{fig:gaugelessZbbinSM}The only diagram contributing to the
SM $G^0 b \bar{b}$ vertex in the gaugeless limit for massless $b$-quarks.}
\end{figure}

In practice, carrying out the calculation in the aforementioned limit
amounts to the elementary computation of the
one-loop off-shell vertex $G^0 b \bar{b}$.
In the gaugeless limit and for massless $b$-quarks the only contribution to
this vertex is depicted in Fig.~\ref{fig:gaugelessZbbinSM}, where the cross
in the top-quark line represents a top-quark mass insertion needed
to flip chirality (an insertion in the other top-quark line is assumed). 
This diagram gives a derivative coupling of the Goldstone
field to the $b$-quarks which can be gauged (or related to the $Z$ vertex
through the Ward identity) to recover the $Z b \bar{b}$ vertex. Then, one
immediately finds
\begin{equation}
\label{eq:deltasm}
\delta g_{L}^{\mathrm{SM}}\approx \frac{\sqrt{2}G_{F}m_{t}^{4}}{(2\pi
)^{4}}\, \int \frac{i d^{4}k}{(k^{2}-m_{t}^{2})^{2}k^{2}}
=\frac{\sqrt{2}G_{F}m_{t}^{2}}{(4\pi )^{2}}~,
\end{equation}
where $G_F$ is the Fermi constant,
and the $m_t^4$ dependence coming from
three Yukawa couplings and one mass insertion is partially compensated by the
$1/m_t^2$ dependence coming from the loop integral.

\begin{figure}[ht]
\includegraphics[scale=0.6]{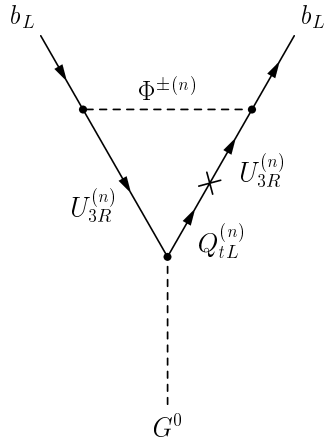}
\caption{\label{fig:gaugelessZbbinUED1}The dominant diagram contributing to the
UED $G^0 b \bar{b}$ vertex in the gaugeless limit for massless $b$-quarks.}
\end{figure}

In the case of a single UED this argument persists: one must simply
consider the analog of diagram in Fig.~\ref{fig:gaugelessZbbinSM}, where now
the particles inside the loop have been replaced by their KK modes,
as shown in Fig.~\ref{fig:gaugelessZbbinUED1}. If we denote by 
$\delta g_L^{UED}$ the new physics contributions in the UED model (the SM
contributions are not included) the result is  
\begin{eqnarray}
\delta g_{L}^{\mathrm{UED}}
&\approx& \frac{\sqrt{2}G_{F}m_{t}^{4}}{(2\pi )^{4}}\,  \sum_{n=1}^\infty
\int
\frac{i d^{4}k}{(k^{2}-m_{Q_t^n}^{2})^{2}(k^{2}-m_{n}^{2})}\nonumber \\ 
&=& \frac{\sqrt{2}G_Fm_t^4}{(4\pi)^2}
\sum_{n=1}^{\infty}\int_0^1
\frac{dx x}{x m_t^2 + m_n^2}
\approx \frac{\sqrt{2}G_{F}m_{t}^{4}}{(4\pi )^{2}} \frac{\pi^2 R^2}{12}~,
\label{eq:deltaUED}
\end{eqnarray}
and depends {\it quartically} on the mass of the top quark. Notice that 
there are several differences with respect to the SM:
(i) The cross now represents the mixing mass term between $Q_t^{(n)}$ and
$U^{(n)}$, which is proportional to $m_t$; 
(ii) The $\Phi^{\pm(n)}$, for $n\not=0$, are essentially 
the physical KK modes of the charged Higgses 
as shown in Eq.\rf{eq:physicalHiggs};
(iii) From the virtual momentum integration one obtains 
now a factor $1/m_n^2$, instead of the  factor $1/m_t^2$ of the SM case. 

This simple calculation allows us to understand easily the leading 
corrections arising from extra dimensions.

A more standard calculation of the $Zb\bar{b}$ 
vertex in UED yields exactly the same result. 
%%%%%%%%%%%%%%%%%%%%%%%%%%%%%%%%%%%%%%%%%%%%%%%%%%%%%%%%%%%%%%%%%%%%%%
\begin{figure}
\begin{displaymath}
\begin{array}{ccc}
\includegraphics[scale=0.8]{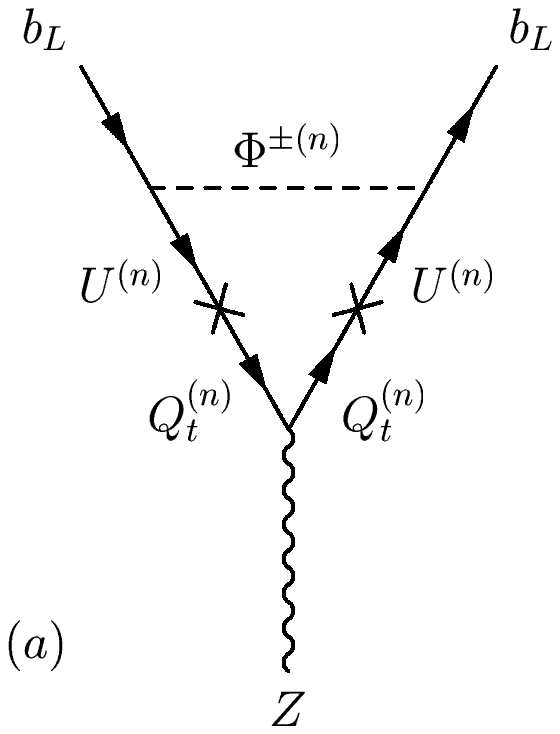} &
\includegraphics[scale=0.8]{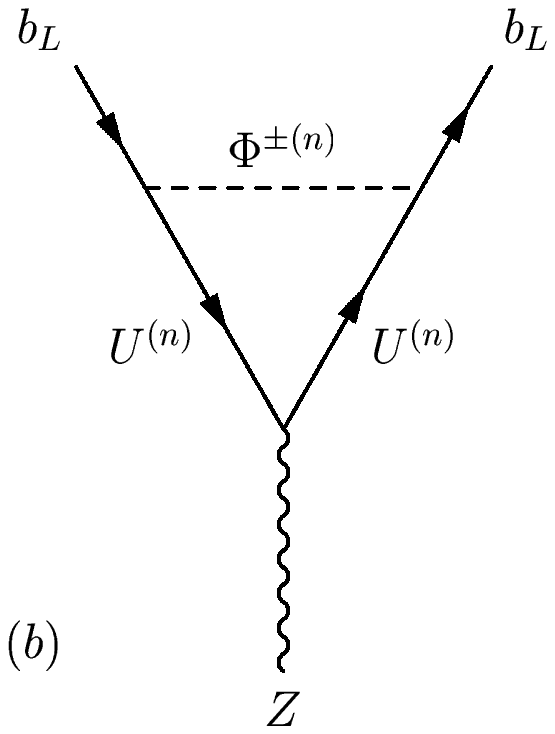} &
\includegraphics[scale=0.8]{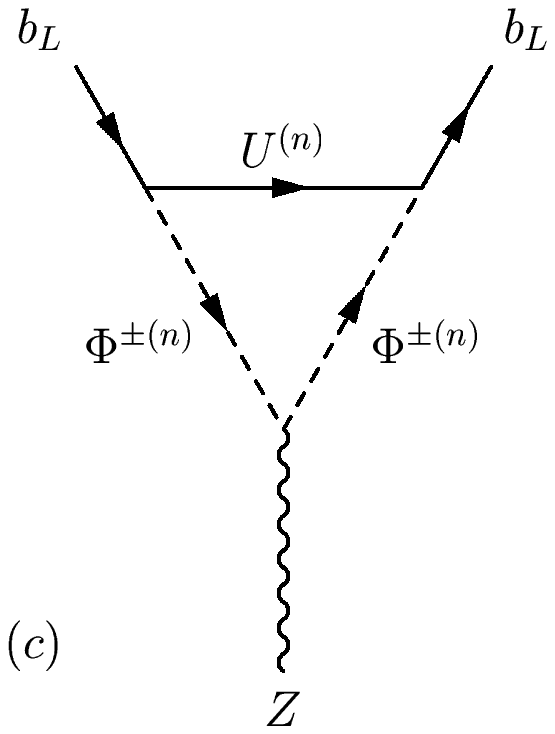}
\end{array}
\end{displaymath}
\begin{displaymath}
\begin{array}{cc}
\includegraphics[scale=0.8]{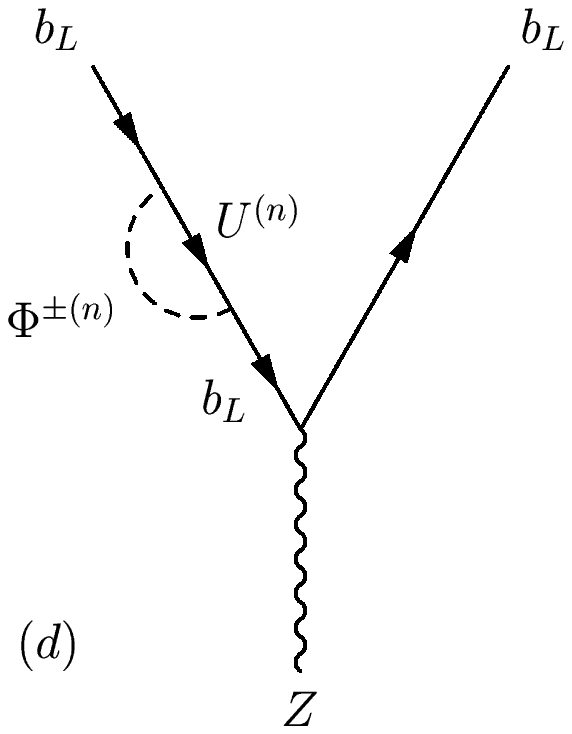} &
\includegraphics[scale=0.8]{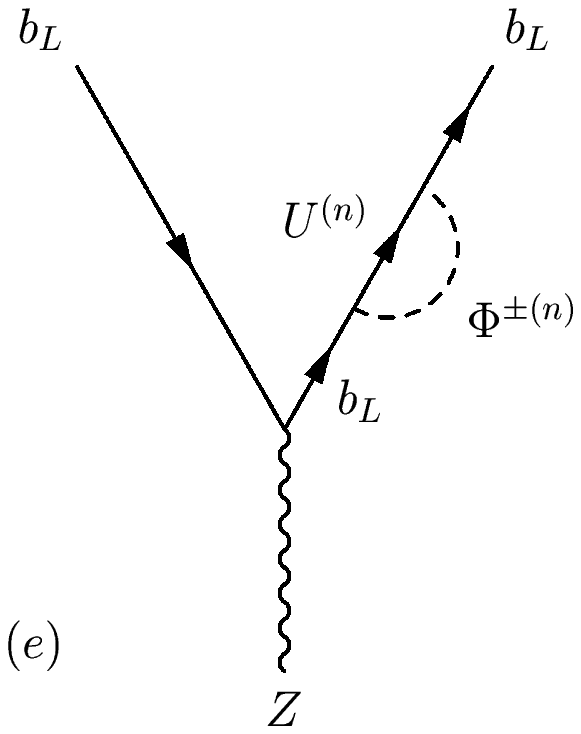}
\end{array}
\end{displaymath}
\caption{Dominant UED contributions to the $Zb\bar{b}$ vertex.}
\label{fig:ZbbinUED}
\end{figure}
%%%%%%%%%%%%%%%%%%%%%%%%%%%%%%%%%%%%%%%%%%%%%%%%%%%%%%%%%%%%%%%%%%%%%%%%%
In this case the radiative corrections to the $Zb\bar{b}$ vertex 
stem from the diagrams of Fig.~\ref{fig:ZbbinUED}.

If we neglect the $b$-quark mass and take $M_Z\ll R^{-1}$, the result, 
for each mode, can be expressed in terms of a single function, $f(r_n)$, 
defined as
\begin{equation}
i {\mathcal M}^{(n)}=
i \frac{g}{c_w} \frac{\sqrt{2}G_F m_t^2}{(4\pi)^2} f(r_n) 
\overline{u}^\prime \gamma^\mu P_L u \epsilon_\mu~,
\end{equation}
where $u$ and $u^\prime$ are the spinors of the $b$ quarks and
$\epsilon_\mu$ stands for the polarization vector of the $Z$
boson. The argument of the function $f$ is $r_n = m_t^2/m_n^2$.

Although the complete result is finite, partial results are divergent and
are regularized by using dimensional regularization.
The contributions of the different diagrams in Fig.~\ref{fig:ZbbinUED} are
\begin{eqnarray}
f^{(a)}(r_n)& =& \left(1-\frac{4}{3}s_w^2 \right)
\left[\frac{r_n-\log(1+r_n)}{r_n}\right]~,\nonumber\\
f^{(b)}(r_n)& = &\left(-\frac{2}{3}s_w^2
\right)\left[\delta_n -1
+
\frac{2r_n+r_n^2-2(1+r_n^2)\log(1+r_n)}{2r_n^2}\right]
 ~,\nonumber\\
f^{(c)}(r_n) & = & \left(-\frac{1}{2}+s_w^2\right) \left[
\delta_n
+ 
\frac{2r_n+3 r_n^2-2(1+r_n)^2 \log(1+r_n)}{2r_n^2}\right]
 ~,\nonumber\\ 
f^{(d)}(r_n)+f^{(e)}(r_n)&=&\left(\frac{1}{2}-\frac{1}{3}s_w^2\right)\left[
\delta_n +
\frac{2r_n+3 r_n^2-2(1+r_n)^2 \log(1+r_n)}{2r_n^2}\right]~,
\label{eq:Zbbresults}
\end{eqnarray}
where 
$
\delta_n \equiv 
 2/\epsilon - \gamma + \log(4\pi) + \log(\mu^2/m_n^2), 
$
and $\mu$ is the 't Hooft mass scale.
From Eq.~(\ref{eq:Zbbresults}) it is straightforward to
verify that all  terms proportional to $ \delta_n$ 
cancel, and so do all  
terms proportional to $s_w^2$, as expected from the gaugeless limit result. 
Thus, finally, 
the only term which survives is the term in $f^{(a)}(r_n)$ 
not proportional to $s_w^2$, yielding the following 
(per mode) contribution to $g_L$: 
\begin{equation}
\delta g_L^{(n)} = \frac{\sqrt{2}G_F m_t^2}{(4\pi)^2} 
\left[\frac{r_n-\log(1+r_n)}{r_n}\right]~,
\end{equation}
which is precisely the one obtained from the gaugeless limit 
calculation, e.g. Eq.~(\ref{eq:deltaUED}) 
with the elementary integration over the Feynman parameter $x$  
already carried out. 
 Notice also that 
the above result is consistent with the decoupling theorem since the
contribution for each mode vanishes when its mass is taken to
infinity, e.g. $r_n\to 0$.

In order to compute the effect of the entire KK tower, it is 
more convenient to first carry out the sum and then 
evaluate the Feynman parameter integral; 
this  interchange is mathematically  
legitimate since the final answer is convergent. 
Thus, 
\begin{equation}
\label{eq:forLUED}
\delta g_L^{\mathrm{UED}} = 
\sum_{n=1}^{\infty}\delta g_L^{(n)} = \frac{\sqrt{2}G_F m_t^2}{(4\pi)^2}
\int_0^1 dx \sum_{n=1}^{\infty}
\frac{r_n x}{1+r_nx} 
= \frac{\sqrt{2}G_F m_t^2}{(4\pi)^2} F_{\mathrm{UED}}(a)~,
\end{equation}
where $a=\pi R m_t$, and 
\begin{eqnarray}
F_{\mathrm{UED}}(a) &=& - \frac{1}{2} + \frac{a}{2} \int_0^1\;dx\;
\sqrt{x}\coth(a\sqrt{x})
\nonumber\\
&\approx& \frac{a^2}{12}-\frac{a^4}{270}+\mathcal{O}(a^6)~.
\label{eq:faued}
\end{eqnarray}

It is instructive to compare the above result with the one  
obtained in the context of models where the extra dimension 
is not universal. In particular, in the 
model considered in  \cite{Papavassiliou:2000pq} the 
fermions live in four dimensions, and only 
the gauge bosons and the Higgs doublet
live in five \cite{Pomarol:1998sd}.  
In this case there is no KK tower for the fermions, and therefore,  
in the loop-diagrams appear only the SM quarks interacting with 
the KK tower of the Higgs fields. 
The result displays a logarithmic dependence on the parameter $a$,
which gives rise to a relatively tight lower bound 
on $R^{-1}$, of the order of 1 TeV. Specifically, 
the corresponding $F(a)$ is given by\footnote{Note that, unlike
in ref.~\cite{Papavassiliou:2000pq}, the $F(a)$ does not include the 
SM contribution.}
\begin{eqnarray}
F(a) &=&-1+2a\int_0^\infty\;dx \frac{x^2}{(1+x^2)^2} \coth(ax)
\nonumber\\
&\approx&
\left(\frac{2}{3}\log(\pi/a)-\frac{1}{3}-\frac{4}{\pi^2}\zeta^\prime(2)\right)
a^2~,
\label{eq:fahg}
\end{eqnarray}
where the expansion on the second line holds for small values of $a$, 
and \( \zeta' \) is the derivative of the Riemann Zeta function. 
The appearance of the $\log(a)$ in $F(a)$ and its absence from 
$F_{\mathrm{UED}}(a)$ may be easily  
understood from the effective theory point of
view. Due to the KK-number conservation in UED models,
the tree-level low energy 
effective Lagrangian when all KK modes are integrated out
is exactly the Standard Model; there are no additional tree-level operators 
suppressed by the compactification scale. Since one-loop logarithmic 
contributions, $\log(a)$, can be obtained in the effective theory by computing 
the running of operators generated at tree level, it is clear that in the 
UED no $\log(a)$ can appear at one loop in low energy observables. The
situation is completely different if higher dimension operators are already
generated at tree level, as is the case of the model considered in 
ref.~\cite{Papavassiliou:2000pq}, where the leading logarithmic corrections
can be computed by using the tree-level effective operators in loops.    

We next turn to the bounds on $R^{-1}$.
We will use the SM prediction 
$R_b^{\mathrm{SM}}=0.21569 \pm 0.00016$ and the experimentally measured value 
$R_b^{\mathrm{exp}}=0.21664 \pm 0.00068$. 
Combining Eq.~\rf{eq:rbab} and Eq.~\rf{eq:forLUED}, we obtain
$F_{\mathrm{UED}}(a)= -0.24 \pm 0.31$, 
and making a weak signal treatment \cite{Feldman:1998qc} we
arrive at the 95\% CL bound $F_{\mathrm{UED}}(a)<0.39$. 
The results for a single UED can be easily
derived from \rf{eq:faued}, yielding
\begin{equation}
R^{-1}> 230\;\mbox{GeV}~. 
\end{equation}
The SM prediction for the left-right asymmetry
$A_b^{SM}=0.9347\pm 0.0001$ and the measured value $A_b^{exp}=0.921\pm
0.020$ gives a looser bound. 

\begin{figure}
\begin{displaymath}
\begin{array}{cc}
\includegraphics[scale=0.6]{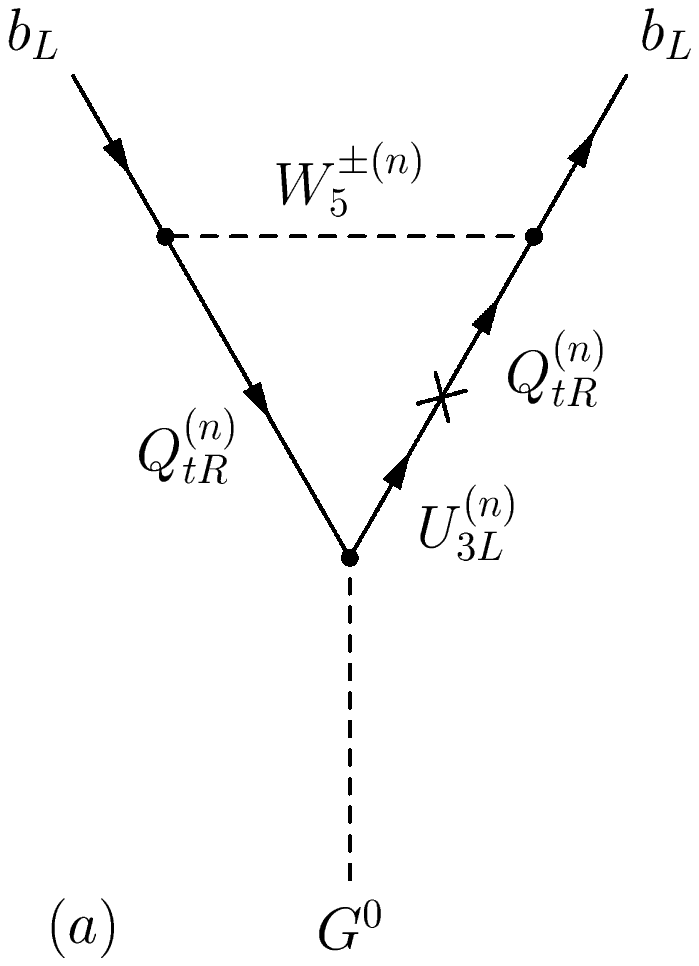}&
\includegraphics[scale=0.6]{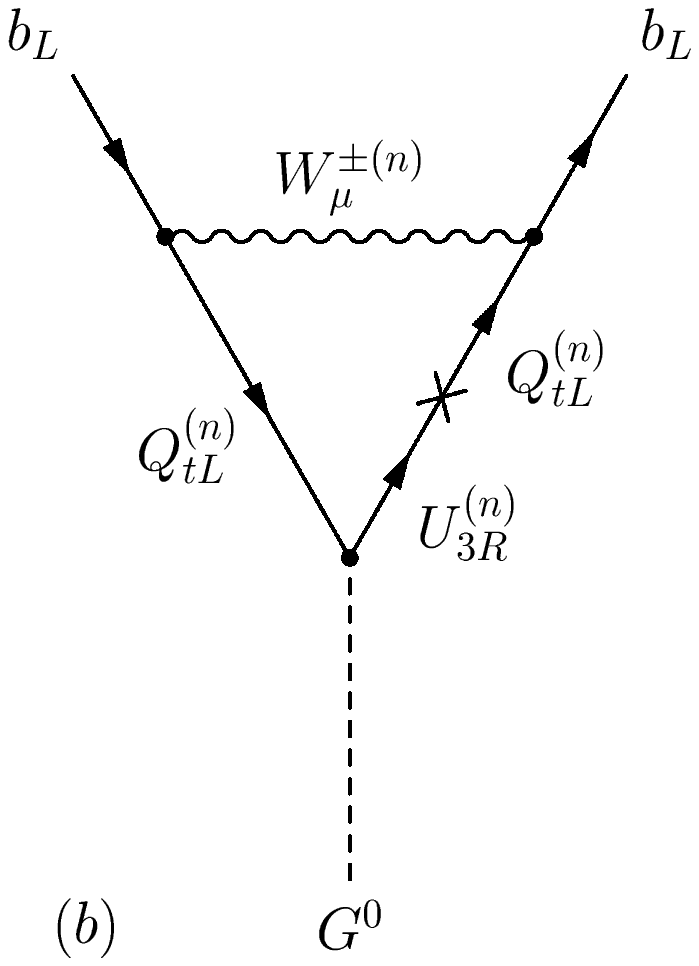}
\end{array}
\end{displaymath}
\caption{Diagrams giving subleading contributions to the $G^{0}b\bar{b}$
vertex.}
\label{fig:subleading}
\end{figure}

Above we have computed only the leading contribution, which goes as $G_F
m_t^4 R^2$. There are also formally subleading contributions, 
suppressed by (at least) an additional factor $M_W^2/m_t^2$; given that
this factor is not so small such corrections could be numerically 
important, and should be estimated. The dominant contributions of
this type come from diagrams with $W_\mu^{\pm(n)}$ and $W_5^{\pm(n)}$ running
in the loops. Since these corrections are still proportional to $m_t^2$ they
can be estimated using again the Ward identity that relates the $G^{0}$
couplings to the $Z$ couplings. 
The relevant diagrams are shown in Fig.~\ref{fig:subleading}.
Their contribution modify the value of $\delta g_L^{\mathrm{UED}}$ 
as follows:
\begin{equation}
\delta g_L^{\mathrm{UED}} = \frac{\sqrt{2}G_F m_t^2}{(4\pi)^2}
F_{\mathrm{UED}}(a) \left(1+3\frac{M_W^2}{m_t^2}\right)~.
\end{equation}
Taking these corrections into account leads to a slight modification 
of the bound on the compactification scale, 
$R^{-1}> 300\;\mbox{GeV}$. 
Evidently, this bound is absolutely comparable to the one obtained from the 
$\rho$ parameter. 

\section{Conclusions}

We have computed the leading contributions, for a large top-quark mass, 
to $Z\rightarrow b\bar{b}$ in a model with one universal extra dimension.
These contributions depend {\em quartically} on the 
top-quark mass and can be evaluated easily in the
gaugeless limit of the theory, where only one diagram contributes. 

There are subleading corrections, formally suppressed by a factor 
$M_W^2/m_t^2$, which, in principle, can be important. 
We have estimated them  by considering the diagrams  with the KK modes of the 
$W^{-(n)}$ and the $W_5^{-(n)}$ running in the loop, and found that they
contribute a $+65\%$ of the correction. 

None of the contributions contains logarithmic ($\log(R)$) corrections. 
This can
be understood from the KK-number conservation, which leads to the absence of
tree-level low-energy operators (containing only SM fields).  
These results have been used to set
a bound on the compactification scale $R^{-1} > 300$~GeV at 95\%~CL
which is comparable to the bounds obtained from the contributions of KK
modes to the $\rho$ parameter \cite{Appelquist:2000nn} in this model, and
which is much weaker than bounds obtained in models with no KK-number
conservation \cite{Papavassiliou:2000pq}. 

What are the consequences of these results for further studies of UED in
b-physics? In \cite{Bernabeu:1997zh} it was shown that the vertex
$Z\rightarrow b \bar{b}$ and $B-\bar{B}$ mixing are highly correlated, and
that it is very difficult to obtain a relatively large contribution to 
$B-\bar{B}$ mixing evading the bounds coming from $R_b$. This was corroborated 
explicitly in \cite{Papavassiliou:2000pq} in a model with only
scalars and gauge bosons in extra dimensions. 
Recently $B-\bar{B}$ mixing has also
been considered in UED~\cite{Chakraverty:2002qk,Buras:2002ej}.
Although the simple argument, developed in \cite{Bernabeu:1997zh}, was
based on the logarithmic corrections and it is
not applicable in the case of UED because of the absence of logarithmic
corrections both in $Z\rightarrow b \bar{b}$ and in $B-\bar{B}$ mixing, 
we believe that some correlation still exists. 
This will be further explored in a future work.

\begin{acknowledgments}
This work has been funded by the MCYT under the Grant BFM2002-00568, 
by the OCYT of the ``Generalitat Valenciana'' under the Grant GV01-94 and by
the CICYT under the Grant AEN-99-0692.
\end{acknowledgments}

%\bibliographystyle{h-physrev3}
%\bibliographystyle{apsrev}
%\bibliography{mybib}

\end{document}